\documentclass{article}
\usepackage{fleqn}
\usepackage{amsmath,amssymb}

\begin{document}
\setcounter{section}{2}
\setcounter{equation}{34}
\begin{center}
{\bf\Large Cosmological evolution  of the cosmological plasma with interpartial scalar interaction.\\[12pt]  II. Formulation of mathematical model.}. \\[12pt]
Yu.G. Ignat'ev\\
N.I. Lobachevsky Institute of Mathematics and Mechanics, Kazan Federal University, \\ Kremleovskaya str., 35, Kazan, 420008, Russia
\end{center}

\begin{abstract} On the basis of the relativistic kinetic theory the relativistic statistical systems with scalar interaction particles are investigated. The self-consistent system of the equations describing self-gravitating plasma with interpartial scalar interaction is formulated, macroscopical laws of preservation are received. The closed system of the equations describing cosmological models to which the matter is presented by plasma with interpartial scalar interaction is received.
\end{abstract}

\section{The Statisical Systems Scalar Interacting Particles}
\subsection{Distribution Function and Macroscopic Flow Densities}
Let $F(x,P)$ is invariant function of distribution of particles in 8-dimensional phase space and let $\psi (x,P)$ is some tensor function of dynamic variables $(x,P)$. According to [1] everyone to tensor dynamic function there corresponds macroscopic density of a flow:\footnote{ There are continuous numbering of formulas on all parts of article, and all formulas till (34), inclusive, are refere to first part of article: Russian Physics Journal
Volume 55, Number 2 (2012), 166-172, DOI: 10.1007/s11182-012-9790-9.
}

\begin{equation} \label{GrindEQ__35_}
\Psi ^{i} (x)=\int _{P(x)} F(x,P)\psi (x,P)\frac{\partial H}{\partial P_{i} } dP\equiv m_{*}^{-1} \int _{P(x)} F(x,P)\psi (x,P)P^{i} dP,
\end{equation}
  Let's define according to \eqref{GrindEQ__35_} moments relativity distribution  $F(x,P)$ [2]:

\begin{equation} \label{GrindEQ__36_}
n^{i} (x)=\int _{P(x)} F(x,P)\frac{\partial H}{\partial P_{i} } dP\equiv m_{*}^{-1} \int _{P(x)} F(x,P)P^{i} dP,
\end{equation}
 is density flow number particles vector\footnote{Numeric vector by J. Singe [3].}, so that:

\begin{equation} \label{GrindEQ__37_}
n^{i} =nv^{i} ,
\end{equation}
 where $v^{i} $-- is the time-like unit vector of kinematical macroscopic velosity of particles:

\begin{equation} \label{GrindEQ__38_}
n=\sqrt{(n,n)} .
\end{equation}
 Further:

\begin{equation} \label{GrindEQ__39_}
T_{p}^{ik} (x)=\int _{P(x)} F(x,P)P^{i} \frac{\partial H}{\partial P_{k} } dP\equiv m_{*}^{-1} \int _{P(x)} F(x,P)P^{i} P^{k} dP,
\end{equation}
 is the macroscopic energy-momentum tensor (EMT). The trace of this tensor can be obtain with take account of the normalization relation \eqref{GrindEQ__8_}:

\begin{equation} \label{GrindEQ__40_}
T_{p} \equiv g_{ik} T_{p}^{ik} =m_{*} \int _{P(x)} F(x,P)dP.
\end{equation}
The invariant volume element of the 4-dimentional momentum space in the expressions \eqref{GrindEQ__36_}, \eqref{GrindEQ__39_} In the system of units accepted by us  is [2]:

\begin{equation} \label{GrindEQ__41_}
dP=\frac{2S+1}{(2\pi )^{3} \sqrt{-g} } dP_{1} \wedge dP_{2} \wedge dP_{3} \wedge dP_{4} ,
\end{equation}
where $S$ is the particle's spin. The invariant 8-dimentional distribution function  $F(x,P)$, which is singularity on the mass-surface \eqref{GrindEQ__8_}, is connected with the nonsingularity 7-dimentional distribution function $f(x,P)$ with help $\delta $-function by the relation [2]:

\begin{equation} \label{GrindEQ__42_}
F(x,P)=f(x,P)\delta (H)=m_{*} \frac{\delta (P_{4} -P_{4}^{+} )}{P_{+}^{4} } ,
\end{equation}
 where $P_{4}^{+} $ is positive root of the normalization equation (8), and $P_{+}^{4} =g^{4k} P_{k}^{+} $ is appropriate this root the value of contrvariant momentum component. In the locally-Lorentzian frame of reference it is:

\begin{equation} \label{GrindEQ__43_}
P_{+}^{4} =\sqrt{m_{*}^{2} +P^{2} } ,
\end{equation}
where $P^{2} =\sum _{\alpha =1}^{3} (P^{\alpha } )^{2} $ is a quard of the physical momentum. Thereby, we obtained the invariant volume element of the 3-dimentional momentum space:

\begin{equation} \label{GrindEQ__44_}
dP_{+} =m_{*} \frac{2S+1}{(2\pi )^{3} \sqrt{-g} } \frac{dP_{1} \wedge dP_{2} \wedge dP_{3} }{P_{+}^{4} } \equiv m_{*} dP_{0} ,
\end{equation}
where

\begin{equation} \label{GrindEQ__45_}
dP_{0} =\sqrt{-g} \frac{2S+1}{(2\pi )^{3} } \frac{dP^{1} dP^{2} dP^{3} }{P_{4}^{+} } \equiv \sqrt{-g} \frac{2S+1}{(2\pi )^{3} } \frac{d^{3} P}{P_{4}^{+} } .
\end{equation}
Then the expressions \eqref{GrindEQ__36_}, \eqref{GrindEQ__39_} and \eqref{GrindEQ__40_} take form (for the simplicity writing we omit summation by sorts of particles):

\begin{equation} \label{GrindEQ__46_}
n^{i} (x)=\frac{2S+1}{(2\pi )^{3} } \int _{P(x)} f(x,P)P^{i} \sqrt{-g} \; \frac{d^{3} P}{P_{4}^{+} } ;
\end{equation}

\begin{equation} \label{GrindEQ__47_}
T_{p}^{ik} (x)=\frac{2S+1}{(2\pi )^{3} } \int _{P(x)} f(x,P)P^{i} P^{k} \sqrt{-g} \; \frac{d^{3} P}{P_{4}^{+} } ;
\end{equation}

\begin{equation} \label{GrindEQ__48_}
T_{p} =\frac{2S+1}{(2\pi )^{3} } m_{*}^{2} \int _{P(x)} f(x,P)\sqrt{-g} \frac{d^{3} P}{P_{4}^{+} } .
\end{equation}
Thereby, the explicit dependence of the macroscopic flows from effective mass $m_{*} $ in the finish expressions is vanish.

\subsection{The General Relativistic Kinetic Equations}

In consequence of a principle of local accordance and the assumption of 4-point-likeness of particles collisions in each act of interpartial interaction the generalised momentum of system of interaction particles iconservates:

\begin{equation} \label{GrindEQ__49_}
\sum _{I} P_{i} =\sum _{F} P'_{i} ,
\end{equation}
where the summation is carried out by all initial, $P_{i} $, and final, $P'_{i} $, states. Let in plasma run the next reactions:

\begin{equation} \label{GrindEQ__50_}
\sum _{A=1}^{m} \nu _{A} a_{A} {\rm \rightleftarrows }\sum _{B=1}^{m'} \nu '_{B} a'_{B} ,
\end{equation}
where $a_{A} $ are symbols of particles, and $\nu _{A} $ its numbers. Thereby, the generalised momentums of initial and final states are equal to:

\begin{equation} \label{GrindEQ__51_}
P_I=\sum\limits_{A=1}^{m}\sum\limits_{\alpha}^{\nu_A}P^\alpha_A, \quad
P_F=\sum\limits_{B=1}^{m'}\sum\limits_{\alpha'}^{\nu_B}P^{\alpha'}_B.
\end{equation}

The distribution functions of partices are defined by the invariant kinetic equations [4]:

\begin{equation} \label{GrindEQ__52_}
[H_{a} ,f_{a} ]=I_{a} (x,P_{a} ),
\end{equation}
where $J_{a} (x,P_{a} )$ is a collision integral:

\begin{equation} \label{GrindEQ__53_}
I_{a} (x,P_{a} )=-\sum  \nu _{a} \int  '_{a} \delta ^{4} (P_{F} -P_{I} )W_{IF} (Z_{IF} -Z_{FI} )\prod _{I,F} 'dP;
\end{equation}

\[W_{FI} =(2\pi )^{4} |M_{IF} |^{2} 2^{-\sum  \nu _{A} +\sum  \nu '_{b} } -\]
is the scattering matrix of  the reaction's chanel  \eqref{GrindEQ__50_}, ($|M_{IF} |$ are the invariant scattering amplitudes);

\[Z_{IF} =\prod _{I} f(P_{A}^{\alpha } )\prod _{F} [1\pm f(P_{B}^{\alpha '} )];\quad Z_{FI} =\prod _{I} [1\pm f(P_{A}^{\alpha } )]\prod _{F} f(P_{B}^{\alpha '} ),\]
the sign «+» corresponds to bosons, and  «-» -- to fermions (details see in [2], [4]).

\subsection{Transport Equations for the Dynamic Values}

The strict consequentions of the general relativity kinetic equations \eqref{GrindEQ__52_} are the transport equations of the dynamic values $\Psi _{a} (x,P_{a} )$ [4]:

\[\nabla _{i} \sum _{a} \int \limits_{P(x)} \Psi _{a} F_{a} \frac{\partial H_{a} }{\partial P_{i} } dP_{a} -\sum _{a} \int\limits_{P(x)} F_{a} [H_{a} ,\Psi _{a} ]dP_{a} =\]

\begin{equation} \label{GrindEQ__54_}
-\!\!\!\sum _{by\; chanels} \int  \left(\sum _{A=1}^{m} \nu _{A} \Psi _{A} -\!\!\sum _{B=1}^{m'} \nu '_{B} \Psi '_{B} \right)\delta ^{4} (P_{F} -P_{I} )(Z_{IF} W_{IF} -Z_{FI} W_{FI} )\prod _{I,F} dP,
\end{equation}
where the summations is carried out by all reactions' chanels \eqref{GrindEQ__67_}.

Believing in \eqref{GrindEQ__54_}, where -- some fundamental charges that persist in reactions \eqref{GrindEQ__67_}, we get the \eqref{GrindEQ__49_}, \eqref{GrindEQ__51_} and \eqref{GrindEQ__67_} equation of plasma number particle flux densities:

\begin{equation} \label{GrindEQ__55_}
\nabla _{i} J_{G}^{i} =0,
\end{equation}
 where:

\begin{equation} \label{GrindEQ__56_}
J_{G}^{i} =\sum _{a} \frac{2S+1}{(2\pi )^{3} } \; g_{a} \int _{P(x)} f(x,P)P^{i} \sqrt{-g} \; \frac{d^{3} P}{P_{4}^{+} }
\end{equation}
is a vector density of fundamental current, corresponding to the some charge $g_{a} $. In particular, the conservation law \eqref{GrindEQ__55_} always takes place for each sorte of particles $b$ ($g_{a} =\delta _{a}^{b} $) under condition of elasticity of their collisions.

Believing in \eqref{GrindEQ__54_} $\Psi _{a} =P^{k} $ we get the (9), \eqref{GrindEQ__49_} and \eqref{GrindEQ__51_} plasma energy-momentum transport equation:

\begin{equation} \label{GrindEQ__57_}
\nabla _{k} T_{p}^{ik} -\sigma \nabla ^{i} \Phi =0,
\end{equation}
 öhere the scalar density of a charge of plasma, $\sigma $, is entered [5]:

\begin{equation} \label{GrindEQ__58_}
\sigma =\frac{1}{2} \sum _{a} \frac{2S+1}{(2\pi )^{3} } \frac{dm_{*}^{2} }{d\Phi } \int _{P(x)} f(x,P)\sqrt{-g} \; \frac{d^{3} P}{P_{4}^{+} } ,
\end{equation}
 In particular, at a choice of a mass function in the form of (27) expression for scalar density of charges becomes:

\begin{equation} \label{GrindEQ__59_}
\sigma =\Phi \sum _{a} \frac{2S+1}{(2\pi )^{3} } q^{2} \int _{P(x)} f(x,P)\sqrt{-g} \; \frac{d^{3} P}{P_{4}^{+} } .
\end{equation}
It is necessary to notice, that the form (EMT) \eqref{GrindEQ__39_} or \eqref{GrindEQ__47_}, and also scalar density of the charge \eqref{GrindEQ__58_}, obtained for scalar charged particles in [4], at given function of Hamilton is an unequivocal consequence of the assumption of preservation of a full momentum in local collisions of particles.

\noindent      In particular, for the system consisting of one-sortable particles, owing to \eqref{GrindEQ__40_} and \eqref{GrindEQ__58_} the relation takes place:

\begin{equation} \label{GrindEQ__60_}
\sigma =\frac{d\ln m_{*} }{d\Phi } T_{p} .
\end{equation}
At a choice of function of effective mass in the form of (27) previous article it is expression becomes simpler and becomes\textit{ fair and for multicomponent system}:

\begin{equation} \label{GrindEQ__61_}
\sigma =\frac{T_{p} }{\Phi } ,
\end{equation}

\subsection{Local thermodynamic equilibrium}

 If time of free movement of particles before collisions, $\tau _{eff} $, much less characteristic time scale of evolution of statistical system, $t$, i.e.,

\begin{equation} \label{GrindEQ__62_}
\tau _{eff} {\rm \ll }t,
\end{equation}
then in statistical system local thermodynamic equilibrium (LTE) is supported. At the conditions of LTE functions of distribution of particles become locally equilibrium form [5]:

\begin{equation} \label{GrindEQ__63_}
f_{a}^{0} (x,P)=\frac{1}{e^{-\gamma _{a} +(\xi ,P_{a} )} \pm 1}
\end{equation}
where $\xi ^{i} (x)$ is time-like vector

\begin{equation} \label{GrindEQ__64_}
\xi ^{2} \equiv (\xi ,\xi )>0,
\end{equation}
moreover at the conditions of a LTE kinematic speed of plasma (see (37)):

\begin{equation} \label{GrindEQ__65_}
v^{i} =\xi ^{i} /\xi                                                                               (65)
\end{equation}
coincides with dynamic speed, and the scalar

\begin{equation} \label{GrindEQ__66_}
\theta (x)=\xi ^{-1}
\end{equation}
becomes local temperature of plasma. Let's underline that circumstance, that at the conditions of a LTE the local temperature $\theta $ and macroscopical speed $v^{i} $ are identical to all a plasma component. Further, in \eqref{GrindEQ__63_} $\gamma _{a} =\gamma _{a} (x)$ is the reduced chemical potential $a$ - plasma components, connected with classical, $\mu _{a} $, by the relation:

\[\gamma _{a} =\frac{\mu _{a} }{\theta } .\]
In the conditions of a LTE chemical potentials of statistical system in which reactions \eqref{GrindEQ__67_} proceed, should satisfy to system of the algebraic equations of chemical equilibrium:

\begin{equation} \label{GrindEQ__67_}
\sum _{A=1}^{m} \nu _{A} \gamma _{A} =\sum _{B=1}^{m'} \nu '_{B} \gamma _{B'} \Leftrightarrow \sum _{A=1}^{m} \nu _{A} \mu _{A} =\sum _{B=1}^{m'} \nu '_{B} \mu _{B'} ,
\end{equation}
 in which all channels of reactions with participation of the given particles should be considered.

\noindent The energy-momentum tensor of particles \eqref{GrindEQ__47_} with respect to the locally equilibrium distribution function \eqref{GrindEQ__63_} accepts structure of a energy-momentum tensor of the ideal liquid:

\begin{equation} \label{GrindEQ__68_}
T_{p}^{ik} =({\rm {\mathcal E}}_{pl} +{\rm {\mathcal P}}_{pl} )v^{i} v^{k} -{\rm {\mathcal P}}_{pl} g^{ik} ,
\end{equation}
where ${\rm {\mathcal E}}_{pl} $ and ${\rm {\mathcal P}}_{pl} $are total density of energy and pressure of plasma.

\noindent For calculation of density \eqref{GrindEQ__46_}, \eqref{GrindEQ__47_} and \eqref{GrindEQ__48_} concerning equilibrium distribution \eqref{GrindEQ__63_} we will pass in locally-Llorentzian frame of reference, moving with a speed $v^{i} $, then we will pass to spherical system of co-ordinates in space of momentums and we will calculate integrals on angular variables.

\noindent Then covariant generalising results and making replacement of momentum variable $p=m_{*} shx$, we will receive for equilibrium scalar density of expression:

\begin{equation} \label{GrindEQ__69_}
n_{a} =\frac{2S+1}{2\pi ^{2} } m_{*}^{3} \int _{0}^{\infty } \frac{sh^{2} xchxdx}{e^{-\gamma _{a} +\lambda _{*} chx} \pm 1} ;
\end{equation}

\begin{equation} \label{GrindEQ__70_}
E_{pl} =\sum _{a} E_{a} =\sum _{a} \frac{2S+1}{2\pi ^{2} } m_{*}^{4} \int _{0}^{\infty } \frac{sh^{2} xch^{2} xdx}{e^{-\gamma _{a} +\lambda _{*} chx} \pm 1} ;
\end{equation}

\begin{equation} \label{GrindEQ__71_}
P_{pl} =\sum _{a} P_{a} =\sum _{a} \frac{2S+1}{6\pi ^{2} } m_{*}^{4} \int _{0}^{\infty } \frac{sh^{4} xdx}{e^{-\gamma _{a} +\lambda _{*} chx} \pm 1} ;
\end{equation}

\begin{equation} \label{GrindEQ__72_}
T_{p} =\sum _{a} \frac{2S+1}{2\pi ^{2} } m_{*}^{2} \int _{0}^{\infty } \frac{sh^{2} xdx}{e^{-\gamma _{a} +\lambda _{*} chx} \pm 1} ;
\end{equation}

\begin{equation} \label{GrindEQ__73_}
\sigma =\sum _{a} \frac{2S+1}{4\pi ^{2} } \frac{dm_{*}^{2} }{d\Phi } m_{*}^{2} \int _{0}^{\infty } \frac{sh^{2} xdx}{e^{-\gamma _{a} +\lambda _{*} chz} \pm 1} ,
\end{equation}
where $\lambda _{*} =m_{*} /\theta $.

\noindent

\section{Self-consistent kinetic model of self-gravitating plasma with interpartial scalar interaction}

 Let's consider the system consisting of multicomponent plasma and classical scalar field$\Phi $.

\subsection{Conformal scalar field}

 Let's consider at first a Lagrange function of classical massive real conformal scalar field $\Phi $ (see, for example, [4,5]):
 \begin{equation}\label{GrindEQ__74_}
 L_{s} =\frac{\varepsilon }{8\pi } \left[g^{ik} \Phi _{,i} \Phi _{,k} +\left(\frac{R}{6} -m_{s}^{2} \right)\Phi ^{2} \right],
\end{equation}
\noindent  where $m_{s} $ is the mass of scalar field quants, the factor $\varepsilon =\pm 1$ considers the character of the scalar interaction ([1], [6]) -- for the systems whith the repulsion of identical scalar charged particles this factor is equal $+1$, and for the systems whith the attraction of identical scalar charged particles this factor is equal $-1$. The Lagrangian of a scalar field resulted here differs multiplier $1/4\pi $ in comparison with a Lagrangian used in [7, 8] (see [1]). This multiplier is entered for maintenance of a canonical form of the equation of a field with a source.

\noindent Besides, in works [7, 8] tensor Richi is received by convolution of tensor Riman on the first and fourth indexes, whereas we use convolution on the first and third index.

\noindent Components of the energy-momentum  tensor  of a scalar field concerning a Lagrange function \eqref{GrindEQ__74_} are equal:

\begin{eqnarray} \label{GrindEQ__75_}
T_{s}^{ik} = &\frac{\varepsilon }{8\pi } \left[\frac{4}{3} \Phi ^{,i} \Phi ^{,k} -\frac{1}{3} g^{ik} \Phi _{,j} \Phi ^{,j} +m_{s}^{2} g^{ik} \Phi ^{2} +\right.\nonumber\\[12pt]
 & + \left.\frac{1}{3} \left(R^{ik} -\frac{1}{2} Rg^{ik} \right)\Phi ^{2} -\frac{2}{3} \Phi \Phi ^{,ik} +\frac{2}{3} g^{ik} \Phi {\rm \square }\Phi \right],
\end{eqnarray}
where

\[{\rm \square }\Phi \equiv g^{ik} \nabla _{i} \nabla _{k} \Phi =\frac{1}{\sqrt{-g} } \frac{\partial }{\partial x^{i} } \sqrt{-g} g^{ik} \frac{\partial }{\partial x^{k} } \Phi \]
is  the d'Alembertian. The covariant divergence from this tensor taking into account commutative relations for the second covariant derivatives of a vector is equal:

\begin{equation} \label{GrindEQ__76_}
\nabla _{k} T_{s}^{ik} =\frac{\varepsilon }{4\pi } \nabla ^{i} \Phi \left({\rm \square }\Phi +m_{s}^{2} \Phi -\frac{R}{6} \Phi \right).
\end{equation}
Einstein's equations for statistical system of scalar charged particles look like:

\begin{equation} \label{GrindEQ__77_}
R^{ik} -\frac{1}{2} Rg^{ik} =8\pi (T_{p}^{ik} +T_{s}^{ik} ),
\end{equation}
where it is necessary to substitute expressions for a component of tensors of energy of a momentum of plasma \eqref{GrindEQ__68_} and a scalar field \eqref{GrindEQ__75_}. Calculating covariant divergences from both parts of the equations of Einstein \eqref{GrindEQ__77_}, we will receive from \eqref{GrindEQ__57_} and \eqref{GrindEQ__76_} the conservation laws summary energy-momentum:

\begin{equation} \label{GrindEQ__78_}
\nabla _{k} (T_{p}^{ik} +T_{s}^{ik} )=\frac{1}{4\pi } \nabla ^{i} \Phi \left[\varepsilon \left({\rm \square }\Phi +m_{s}^{2} \Phi -\frac{R}{6} \Phi \right)+4\pi \sigma \right]=0,
\end{equation}
whence taking into account $\Phi \not \equiv {\kern 1pt} Const{\kern 1pt} $ we will receive the equation of a massive scalar field with a source (see [4]):

\begin{equation} \label{GrindEQ__79_}
{\rm \square }\Phi +m_{s}^{2} \Phi -\frac{R}{6} \Phi =-4\pi \varepsilon \sigma .
\end{equation}
System of the equations \eqref{GrindEQ__52_}, \eqref{GrindEQ__57_}, \eqref{GrindEQ__77_} and \eqref{GrindEQ__79_} together with definitions \eqref{GrindEQ__47_} and \eqref{GrindEQ__58_} also represent the required closed system of the self-consistent equations describing statistical system of particles with scalar interaction. Calculating a trace of a tensor of energy-impulse of a scalar field \eqref{GrindEQ__75_}, we will receive taking into account \eqref{GrindEQ__79_}:

\begin{equation} \label{GrindEQ__80_}
T_{s} =\frac{\varepsilon }{4\pi } m_{s}^{2} \Phi ^{2} -\varepsilon \sigma \Phi .
\end{equation}
 In particular, at a choice of a mass function in the form of (27) taking into account \eqref{GrindEQ__61_} we will receive from here:
 \begin{equation}\label{GrindEQ__81_}
 T=T_{p} +T_{s} =\frac{\varepsilon }{4\pi } m_{s}^{2} \Phi ^{2} .
 \end{equation}

\subsection{Nonconformal scalar field}

In this case the Lagrangian of a real scalar field can be chosen in a form:

\begin{equation} \label{GrindEQ__82_}
L_{s} =\frac{\varepsilon }{8\pi } \left(g^{ik} \Phi _{,i} \Phi _{,k} -m_{s}^{2} \Phi ^{2} \right).
\end{equation}
 Then the energy-momentum tensor of a scalar field is:

\begin{equation} \label{GrindEQ__83_}
T_{s}^{ik} =\frac{\varepsilon }{8\pi } \left(2\Phi ^{,i} \Phi ^{,k} -g^{ik} \Phi ^{,j} \Phi _{,j} +g^{ik} m_{s}^{2} \Phi ^{2} \right).
\end{equation}
 Similarly \eqref{GrindEQ__78_} we will receive the conservation law:

\begin{equation} \label{GrindEQ__84_}
\nabla _{k} (T_{p}^{ik} +T_{s}^{ik} )=\frac{1}{4\pi } \nabla ^{i} \Phi \left[\varepsilon ({\rm \square }\Phi +m_{s}^{2} \Phi )+4\pi \sigma \right]=0,
\end{equation}
 whence we will obtaine the equation of a massive nonconformal scalar field with a source (see [4]):

\begin{equation} \label{GrindEQ__85_}
{\rm \square }\Phi +m_{s}^{2} \Phi =-4\pi \varepsilon \sigma .
\end{equation}
The trace of an energy-momentum tensor of a scalar field \eqref{GrindEQ__83_} is equal:

\begin{equation} \label{GrindEQ__86_}
T_{s} =\frac{\varepsilon }{4\pi } (-\Phi ^{,j} \Phi _{,j} +2m_{s}^{2} \Phi ^{2} ).
\end{equation}

\section{Self-consistent cosmological model for the local equilibrium plasma with interpartial scalar interaction }

In case of performance of condition LTE \eqref{GrindEQ__62_}  the collisions integral in the right part of the kinetic equations becomes the large values, therefore for locally equilibrium plasma instead of the solving of the kinetic equations it is necessary to take advantage of definition of a tensor of energy-momentum of a liquid \eqref{GrindEQ__68_}, and also relations \eqref{GrindEQ__69_} - \eqref{GrindEQ__73_}, defining macroscopical scalars, and the equations of chemical balance \eqref{GrindEQ__67_}. Thus it is necessary to consider, that locally-equilibrium distribution functions \eqref{GrindEQ__63_} at performance of conditions of chemical balance automatically turn into zero integral of collisions \eqref{GrindEQ__53_}.

However, it agree to the logician of hydrodynamic approximation (see, for example, [1]) equality to zero of the right part of the kinetic equations in this case is necessary for understanding only as the approached parity, fair only for the macroscopical moments of locally equilibrium function of distribution.

\subsection{Self-consistent system of the equations for isotropic homogeneous spatially flat Universe}

Let's consider spatially-flat cosmological model of the Freedman

\begin{equation} \label{GrindEQ__87_}
ds^{2} =dt^{2} -a^{2} (t)(dx^{2} +dy^{2} +dz^{2} ),
\end{equation}
when the matter consists from the equilibrium plasma of scalar intetacting particles and the massive scalar field, depending only from cosmological time, $\Phi (t)$.  To condition of rest of plasma concerning synchronous in the metrics \eqref{GrindEQ__87_} frame of reference there corresponds a vector of macroscopical speed:

\begin{equation} \label{GrindEQ__88_}
v^{i} =\delta _{4}^{i} .
\end{equation}
Components of the Einstein tensor of concerning the metrics \eqref{GrindEQ__87_} are equal:

\begin{equation} \label{GrindEQ__89_}
G_{k}^{i} =2\frac{\dot{a}^{2} -a\ddot{a}}{a^{2} } v^{i} v_{k} +\frac{\dot{a}^{2} +2a\ddot{a}}{a^{2} } \delta _{k}^{i} .
\end{equation}
Further, calculating components, $\Phi _{\; ,k}^{,i} $ we will find:

\begin{equation} \label{GrindEQ__90_}
\Phi _{\; ,k}^{,i} =\left(\ddot{\Phi }-\frac{\dot{a}}{a} \dot{\Phi }\right)v^{i} v_{k} +\frac{\dot{a}}{a} \dot{\Phi }\delta _{k}^{i} .
\end{equation}
From \eqref{GrindEQ__89_} and \eqref{GrindEQ__80_} follows:

\begin{equation} \label{GrindEQ__91_}
R=-6\frac{\dot{a}^{2} +a\ddot{a}}{a^{2} } =-8\pi \varepsilon \sigma \Phi +2\varepsilon m_{s}^{2} \Phi ^{2} .
\end{equation}

\noindent \textit{Conformal scalar field. }The equation of the confomal scalar field \eqref{GrindEQ__79_} in the metrics \eqref{GrindEQ__87_} taking into account \eqref{GrindEQ__91_} can be written down in a kind:

\begin{equation} \label{GrindEQ__92_}
\ddot{\Phi }+3\frac{\dot{a}}{a} \dot{\Phi }+m_{s}^{2} \Phi \left(1+\frac{1}{3} \Phi ^{2} \right)=-4\pi \varepsilon \sigma (1-\Phi ^{2} ).
\end{equation}
Considering parities \eqref{GrindEQ__89_} - \eqref{GrindEQ__91_}, we will calculate components of a tensor of energy-momentum of a confomal scalar field \eqref{GrindEQ__75_} and we will present them in a form a components of a tensor of energy-impulse of an ideal liquid

\noindent

\begin{equation} \label{GrindEQ__93_}
\mathop{T}\limits_{s} {\rm \; }_{k}^{i} =(E_{s} +P_{s} )v^{i} v_{k} -P_{s} \delta _{k}^{i} ,
\end{equation}
 where $E_{s} $ è $P_{s} $  are energy density and pressure of the scalar field, respectively:

\begin{equation} \label{GrindEQ__94_}
E_{s} =\frac{\varepsilon }{8\pi } \left[\left(\frac{1}{a} \frac{d}{dt} a\Phi \right)^{2} +m_{s}^{2} \Phi ^{2} \right];
\end{equation}

\begin{equation} \label{GrindEQ__95_}
P_{s} =\frac{\varepsilon }{24\pi } \left[\left(\frac{1}{a} \frac{d}{dt} a\Phi \right)^{2} -m_{s}^{2} \Phi ^{2} +8\pi \sigma \Phi \right].
\end{equation}
It is easy to verify up, that in consequence of \eqref{GrindEQ__94_} and \eqref{GrindEQ__95_} the parity \eqref{GrindEQ__80_} for a trace of a tensor of energy-momentum of a scalar field is identically carried out.

\noindent

\textit{ Nonconformal scalar field..} In this case we will obtain from \eqref{GrindEQ__83_}:

\begin{equation} \label{GrindEQ__96_}
E_{s} =\frac{\varepsilon }{8\pi } (\dot{\Phi }^{2} +m_{s}^{2} \Phi ^{2} );\quad P_{s} =\frac{\varepsilon }{8\pi } (\dot{\Phi }^{2} -m_{s}^{2} \Phi ^{2} ).
\end{equation}
Thus for a trace of a tensor of energy-momentum of a scalar field we will obtain:

\begin{equation} \label{GrindEQ__97_}
T_{s} =E_{s} -3P_{s} =\frac{\varepsilon }{4\pi } (-\dot{\Phi }^{2} +2m_{s}^{2} \Phi ^{2} ),
\end{equation}
 and also:

\begin{equation} \label{GrindEQ__98_}
E_{s} +P_{s} =\varepsilon \frac{\dot{\Phi }^{2} }{4\pi } .
\end{equation}

\subsection{Complect system of equentions for the local equilibrium plasma}

 Let's formulate now the closed system of the equations, defining cosmological evolution of locally equilibrium plasma with interpartial scalar interaction. As is known, in the metrics Einstein's \eqref{GrindEQ__87_} independent equations become:

\begin{equation} \label{GrindEQ__99_}
3\frac{\dot{a}^{2} }{a^{2} } =8\pi E;
\end{equation}

\begin{equation} \label{GrindEQ__100_}
\dot{E}+3\frac{\dot{a}}{a} (E+P)=0.
\end{equation}
The equation of the transport  energy-momentum of a plasma \eqref{GrindEQ__57_} in the metrics \eqref{GrindEQ__87_} can be written down in a form:

\begin{equation} \label{GrindEQ__101_}
\dot{E}_{pl} +3\frac{\dot{a}}{a} (E_{pl} +P_{pl} )=\sigma \dot{\Phi }.
\end{equation}
Let further locally equilibrium plasma consists from $N$ various sorts of particles $a=\overline{1,N}$. Then all considered cosmological system "plasma+scalar field"» is defined by  $N+3$ functions of time $a(t),\Phi (t),\gamma _{a} (t),\theta (t)$, where $\theta $ - local temperature of plasma, $\gamma _{a} $ - the reduced chemical potentials, defining locally-equilibrium functions of distribution \eqref{GrindEQ__63_}. Chemical potentials $\gamma _{a} $ are defined from the conditions of chemical balances \eqref{GrindEQ__67_}, representing system of the linear homogeneous algebraic equations and inheriting symmetry of algebra of interaction of particles of plasma. As a result of homogeneity of equations system of the  chemical balance, at least, one of functions $\gamma _{c} (t)$ remains uncertain. This function should be found from the  conservation law of some fundamental charge $G$ \eqref{GrindEQ__55_}. In a cosmological situation this law looks like:

\begin{equation} \label{GrindEQ__102_}
a^{3} (t)\sum _{a=1}^{N} g_{a} \gamma _{a} (t)={\kern 1pt} Const{\kern 1pt}
\end{equation}
The remained three unknown functions, $\theta (t),a(t),\Phi (t)$ are defined from the equations \eqref{GrindEQ__99_} - \eqref{GrindEQ__101_}, and it is necessary to notice, that the equation \eqref{GrindEQ__100_}, in turn, is a consequence of the equation of a scalar field ((79) or \eqref{GrindEQ__85_}) and the transport equations  over \eqref{GrindEQ__101_}. Therefore further we will replace with its corresponding equation of a field.

\noindent

Thus, the equations \eqref{GrindEQ__99_}, \eqref{GrindEQ__101_}, \eqref{GrindEQ__79_} (or (85)) together with the equations of chemical balance \eqref{GrindEQ__67_} and conservation laws of fundamental charges \eqref{GrindEQ__102_} make full self-consistent system of the equations, defining evolution of locally equilibrium plasma with interpartial scalar interaction.

In following articles we will consider the numerical models of the cosmological expansion, based on system presented here of the equations.

\noindent E-mail: ignatev\_yu@rambler.ru

\end{document}